\documentclass[aip,jap,reprint]{revtex4-1}
\usepackage[T1]{fontenc}
\usepackage[latin9]{inputenc}
\usepackage{amstext}
\usepackage{graphicx}
\usepackage{amssymb}
\usepackage{multirow}
\usepackage{amsmath,amssymb}


\begin{document}

\preprint{}

\title{Sputtered TiN films for superconducting coplanar waveguide resonators}

\author{S. Ohya}
\email[Current address: Department of Electrical Engineering and Information Systems, The University of Tokyo, 7-3-1 Hongo, Bunkyo-ku, Tokyo 113-8656, Japan. Electronic mail: ]{ohya@cryst.t.u-tokyo.ac.jp}
\affiliation{Department of Electrical and Computer Engineering, University of California, Santa Barbara, California 93106-9560, USA}
\author{B. Chiaro}
\affiliation{Department of Physics, University of California, Santa Barbara, California 93106, USA}
\author{A. Megrant}
\affiliation{Department of Physics, University of California, Santa Barbara, California 93106, USA}
\affiliation{Department of Materials, University of California, Santa Barbara, California 93106-9530, USA}
\author{C. Neill}
\affiliation{Department of Physics, University of California, Santa Barbara, California 93106, USA}
\author{R. Barends}
\affiliation{Department of Physics, University of California, Santa Barbara, California 93106, USA}
\author{Y. Chen}
\affiliation{Department of Physics, University of California, Santa Barbara, California 93106, USA}
\author{J. Kelly}
\affiliation{Department of Physics, University of California, Santa Barbara, California 93106, USA}
\author{D. Low}
\affiliation{Department of Physics, University of California, Santa Barbara, California 93106, USA}
\author{J. Mutus}
\affiliation{Department of Physics, University of California, Santa Barbara, California 93106, USA}
\author{P. J. J. O'Malley}
\affiliation{Department of Physics, University of California, Santa Barbara, California 93106, USA}
\author{P. Roushan}
\affiliation{Department of Physics, University of California, Santa Barbara, California 93106, USA}
\author{D. Sank}
\affiliation{Department of Physics, University of California, Santa Barbara, California 93106, USA}
\author{A. Vainsencher}
\affiliation{Department of Physics, University of California, Santa Barbara, California 93106, USA}
\author{J. Wenner}
\affiliation{Department of Physics, University of California, Santa Barbara, California 93106, USA}
\author{T. C. White}
\affiliation{Department of Physics, University of California, Santa Barbara, California 93106, USA}
\author{Y. Yin}
\affiliation{Department of Physics, University of California, Santa Barbara, California 93106, USA}
\author{B. D. Schultz}
\affiliation{Department of Electrical and Computer Engineering, University of California, Santa Barbara, California 93106-9560, USA}
\author{C. J. Palmstr\o m}
\affiliation{Department of Electrical and Computer Engineering, University of California, Santa Barbara, California 93106-9560, USA}
\affiliation{Department of Materials, University of California, Santa Barbara, California 93106-9530, USA}
\author{B. A. Mazin}
\affiliation{Department of Physics, University of California, Santa Barbara, California 93106, USA}
\author{A. N. Cleland}
\affiliation{Department of Physics, University of California, Santa Barbara, California 93106, USA}
\author{John M. Martinis}
\email{martinis@physics.ucsb.edu}
\affiliation{Department of Physics, University of California, Santa Barbara, California 93106, USA}

\date{\today}
\begin{abstract}
We present a systematic study of the properties of TiN films by varying the deposition conditions in an ultra-high-vacuum reactive magnetron sputtering chamber. By increasing the deposition pressure from 2 to 9 mTorr while keeping a nearly stoichiometric composition of Ti$_{1-x}$N$_{x}$ ($x$=0.5), the film resistivity increases, the dominant crystal orientation changes from (100) to (111), grain boundaries become clearer, and the strong compressive in-plane strain changes to weak tensile in-plane strain. The TiN films absorb a high concentration of contaminants including hydrogen, carbon, and oxygen when they are exposed to air after deposition.  With the target-substrate distance set to 88 mm the contaminant levels increase from $\sim 0.1\%$ to $\sim 10 \%$ as the pressure is increased from 2 to 9 mTorr.  The contaminant concentrations also correlate with in-plane distance from the center of the substrate and increase by roughly two orders of magnitude as the target-substrate distance is increased from 88 mm to 266 mm.  These contaminants are found to strongly influence the properties of TiN thin films.  For instance, the resistivity of stoichiometric films increases by around a factor of 5 as the oxygen content increases from $0.1\%$ to $11\%$.  These results strongly suggest that the energy of the sputtered TiN particles plays a crucial role in determining the TiN film properties, and that it is important to precisely control the energy of these particles to obtain high-quality TiN films.  Superconducting coplanar waveguide resonators made from a series of nearly stoichiometric films grown at pressures from 2 mTorr to 7 mTorr show a substantial increase in intrinsic quality factor from $\sim 10^{4}$ to $\sim 10^{6}$ as the magnitude of the compressive strain decreases from nearly 3800 MPa to approximately 150 MPa and the oxygen content increases from 0.1$ \%$ to 8$\%$.  The films with a higher oxygen content exhibit lower loss, but the nonuniformity of the oxygen incorporation, which presents as a radially dependent resistivity, hinders the use of reactively sputtered TiN in larger circuits.
\end{abstract}
\maketitle

\section{INTRODUCTION}

Superconducting coplanar-waveguide (SCPW) resonators are used for photon detection and quantum information processing. Recently, there has been a growing interest in titanium nitride (TiN) thin films due to their widely tunable critical temperature $T_{\text{c}}$, large surface inductance, and ability to produce high intrinsic quality-factor $Q_{i}$ resonators.\cite{Leduc2010, Vissers2010, Sage2011, Vissers2012, Diener2012, Mazin2012, Cecil2012, Noroozian2012, Calvo2012, Eom2012, Krocknberger2012, Driessen2012} Although excellent performance has been achieved with TiN SCPW resonators, their loss mechanisms are still not clear due to the complex properties of TiN.  TiN films are known to absorb contaminants when they are exposed to air.\cite{Kumar1988, Mandl1990,Logothetidis1999} In fact, a high concentration of oxygen, up to $\sim$20\%, has been reported in TiN films.\cite{Chowdhury1996, Chappe2007, Radecka2011} Since the contaminants absorbed from the air are strong candidates for two-level systems (TLSs) that can cause degradation in the performance of superconducting devices,\cite{Martinis2005}a systematic investigation of the film quality of TiN is quite important. Although many studies have been done on TiN since the 1980's, it remains difficult to relate sputtering conditions to the properties of the resulting films and there is little information linking the film properties to the performance of microwave electronic devices made from the films. Here, we show a detailed analysis focusing on stoichiometric Ti$_{1-x}$N$_{x}$ films ($x$=0.5) obtained by adjusting the N$_{2}$ flow rate.  We show that the kinetic energy of the sputtered TiN particles, which is a function of the pressure, radial position, and target-substrate (T-S) distance, plays a crucial role in determining film properties such as strain, resistivity, grain structure, crystallographic texture, and contaminant levels.  We find that the quality factors of resonators made from TiN depend strongly on the material properties of the thin films.  Specifically, we find the low power Q$_{i}$ to be correlated with reduced film strain and increased O content.  We find that resonators with a low power Q$_{i} > 10^{6}$ can be reliably produced from low film strain TiN.

\section{FILM PREPARATION}

TiN films were deposited by DC reactive magnetron sputtering in an ultra-high vacuum deposition chamber (AJA International, Inc.) with a background pressure in the lower  $10^{-10}$ Torr range. This system has a high-vacuum load lock chamber connected to the main deposition chamber. We used a 6-inch gun with a 99.995\% purity, 4-inch Ti target. The substrate and target face one another and are centered on a common axis. Ultra-high purity (99.9999\%) Ar and N$_{2}$ gas sources were introduced to the deposition chamber through Micro Torr purifiers (SAES Pure Gas, Inc.). All depositions were done at room temperature with a fixed Ar flow rate of 15 sccm and with a constant DC plasma power of 600 W.  Under these conditions the nominal incident energy of Ar$^{+}$ ions on the Ti target is from 350 to 380 eV, depending on the deposition pressure and N$_{2}$ flow rate.  During deposition, the substrate holder was rotated at $\sim$30 rpm. No substrate bias was applied. We used high-resistivity Si(001) substrates (>10,000 $\Omega$ cm, Addison Engineering, Inc.) for all depositions. Before installing a Si substrate in the load lock, the substrate was cleaned with Nano-Strip (Cyantek Corp, Inc.) for 10 minutes, followed by buffered-HF cleaning for 1 minute to remove any native oxide and terminate the Si surface with hydrogen. We installed the Si substrate in the load lock as quickly as possible after cleaning, typically within 45 minutes. Before deposition of TiN, the Ti target was pre-sputtered for 2 minutes with the same conditions as for the subsequent TiN deposition. In sections \ref{BASIC PROPERTIES} to \ref{IN-PLANE DISTRIBUTION} of this paper, we used a fixed target-substrate (T-S) distance of 88 mm, whereas in section \ref{T-S DISTANCE DEPENDENCE}, we investigated the effect of the T-S distance on the properties of the TiN films.  Film thicknesses varied from 100 to 900 nm.

\section{BASIC PROPERTIES}
\label{BASIC PROPERTIES}

\begin{figure}
\begin{center}
\includegraphics{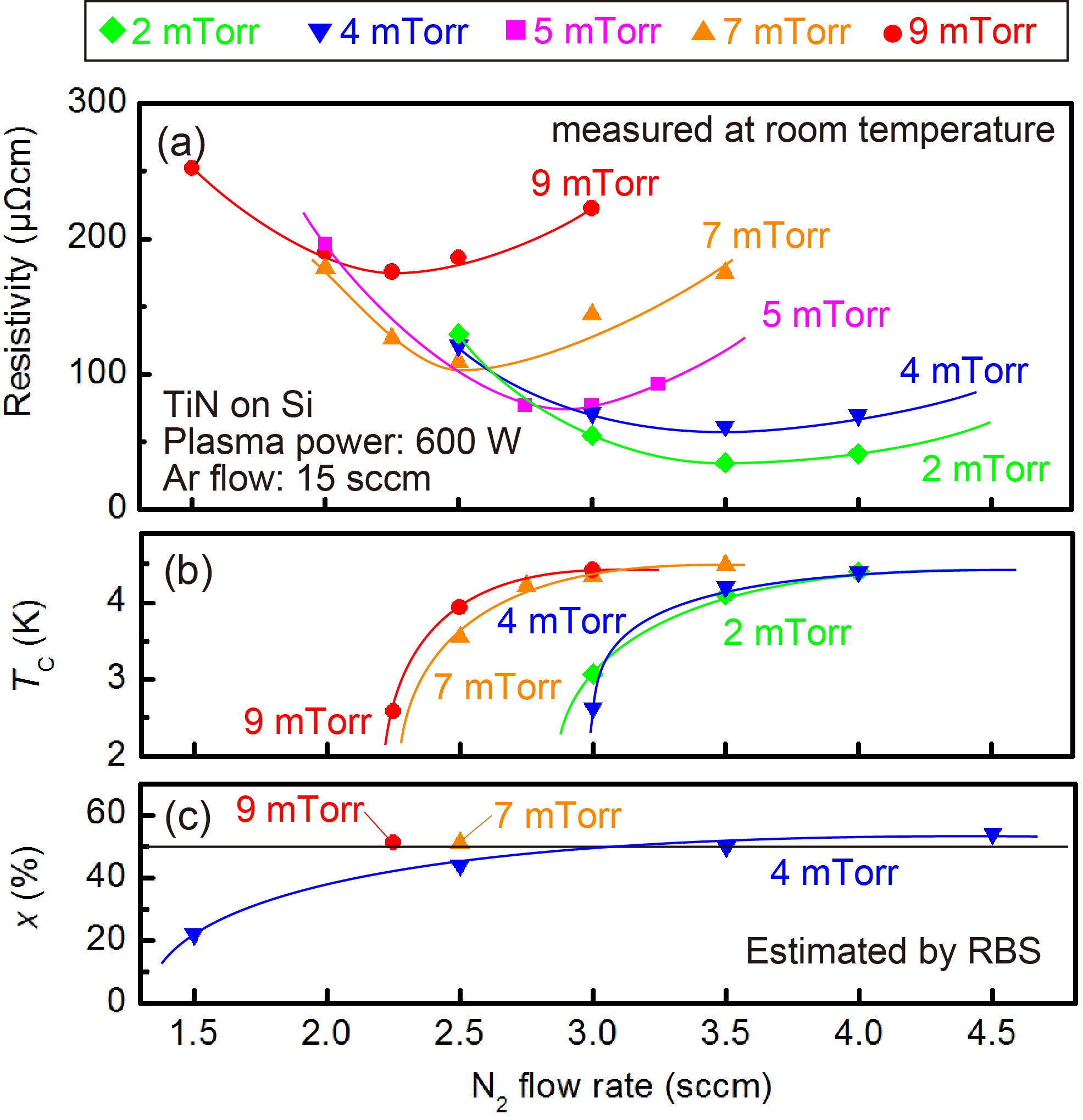}
\end{center}
\caption{(Color) N$_{2}$ flow-rate dependence of the (a) room-temperature resistivity, (b) $T_{\text{c}}$, and (c) $x$ of Ti$_{1-x}$N$_{x}$ films deposited at 2, 4, 5, 7, and 9 mTorr, where we estimated $x$ by using RBS. The curves are guides for eyes. All of these depositions were done at room temperature with the Ar flow rate fixed at 15 sccm and the plasma power of 600 W. For the measurements of the resistivity and $T_{\text{c}}$, the film thickness was $\sim$200 nm for the TiN films deposited at 2 mTorr and was $\sim$100 nm for others.}
\label{BasicProperties}
\end{figure}

Figure\,\ref{BasicProperties}(a)-(c) shows the N$_{2}$ flow-rate dependence of the TiN thin film room temperature resistivity, $T_{\text{c}}$, and composition $x$, defined as the N content divided by the sum of the Ti and N contents. The film thickness,  $T_{\text{c}}$, and $x$ were measured by scanning electron microscopy (SEM), a Physical Property Measurement System (Quantum Design, Inc.), and Rutherford Back Scattering (RBS), respectively. Using RBS we can determine the film composition with an error of 1-2\%. In the RBS measurements, we detected Ti, N, C, and O signals. We did not see any clear Ar signal, which was previously reported in sputtered TiN films.\cite{Williams1987}  Resistivity values were measured within several hours of deposition.  Although a gradual evolution in the film oxygen content has been reported over the timescale of 100 hours after deposition\cite{Logothetidis1999}, our films only show 7\% increase in resistivity one month after deposition, which means that our films are relatively stable.  We measured $T_{\text{c}}$ and $x$ several weeks after deposition. Here, the films were deposited on quarter pieces of 3-inch Si wafers.  All measurements were done near the center of these pieces.

Increasing the N$_{2}$ flow rate increases the N concentration on the target surface, which increases $x$. By comparing Fig.\,\ref{BasicProperties}(a) with (c), we see that the resistivity minimum corresponds to the stoichiometric condition ($x$=0.5) regardless of deposition pressure. This is consistent with previous studies of TiN carried out at a relatively low deposition pressure.\cite{Ahn1983,Schiller1984}   Our results show that the resistivity minimum of the resistance vs. N$_{2}$ flow at constant pressure curve is a good indicator of the stoichiometric point even with a high contamination density.  This is important because in section \ref{PROPERTIES OF STOICHIOMETRIC FILMS} it will be shown that resistivity is a strong function of O content which varies from $\sim$0.1\% to $\sim$10\% over this pressure range. We note that the lowest resistivity film obtained in this study was 34.2 $\mu$$\Omega$ cm. This is a typical value obtained for TiN films deposited at room temperature and slightly higher than single-crystal TiN (18 $\mu$$\Omega$ cm).\cite{Johansson1985} In Fig.\,\ref{BasicProperties}(b), $T_{\text{c}}$ increases and saturates at 4.5 K while increasing the N$_{2}$ flow rate. As $x$ increases in Ti$_{1-x}$N$_{x}$, the dominant phase of the film is changed from Ti$_{2}$N, whose $T_{\text{c}}$ is 50 mK, to TiN, with $T_{\text{c}}$ around 5.6 K.\cite{Wriedt1987, Vissers2013, Penilla2008} This phase change can explain the steep change of $T_{\text{c}}$ when $x$<0.5. In Fig.\,\ref{BasicProperties}(b), we see that the $T_{\text{c}}$ of the stoichiometric TiN film decreases by increasing the deposition pressure, which we believe is due to the increase in contaminant concentrations discussed in section \ref{PROPERTIES OF STOICHIOMETRIC FILMS} (See Fig. \ref{BasicProperties} with the N$_{2}$ flow rate at 3.5, 3.5, 2.5, and 2.25 sccm when the deposition pressure is 2, 4, 7, and 9 mTorr, respectively).

In Fig.\,\ref{BasicProperties}(a), the resistivity rises substantially when increasing the deposition pressure, and the stoichiometric point shifts to a smaller N$_{2}$ flow rate. The resistivity increase can be explained by the morphology and contamination changes discussed below. The shift of the stoichiometric point is probably due to the effect of the Ar neutrals reflected from the Ti target to the substrate during sputtering. These neutrals have the same order of mean free path as the sputtered particles,\cite{Lu2012} and re-sputter the TiN film surface during the deposition. It is known that the film surface is always covered with a stable N-rich TiN thin layer due to the high reactivity of atomic N, which has an important role in determining the N content of the film. When the reflected Ar neutrals have a high energy at a low deposition pressure, re-sputtering by the Ar neutrals removes this N-rich surface. However, at high pressure, this effect becomes less important due to the higher collision probability of the Ar neutrals, so stoichiometric TiN is obtained at a smaller N$_{2}$ flow rate as the deposition pressure increases.\cite{Adjaottor1995}

\section{PROPERTIES OF STOICHIOMETRIC FILMS}
\label{PROPERTIES OF STOICHIOMETRIC FILMS}

In this section, we focus on nearly stoichiometric Ti$_{1-x}$N$_{x}$ films ($x \simeq 0.5$).  Films were deposited by setting the N$_{2}$ flow rate at 3.5, 3.5, 3.0, 2.5, and 2.25 sccm when the deposition pressure was 2, 4, 5, 7, and 9 mTorr, respectively. These N$_{2}$ flow rates correspond to the resistivity minimum points in Fig.\,\ref{BasicProperties}(a). We fixed the Ar flow rate at 15 sccm and the T-S distance at 88 mm.

\begin{figure}
\begin{center}

\includegraphics{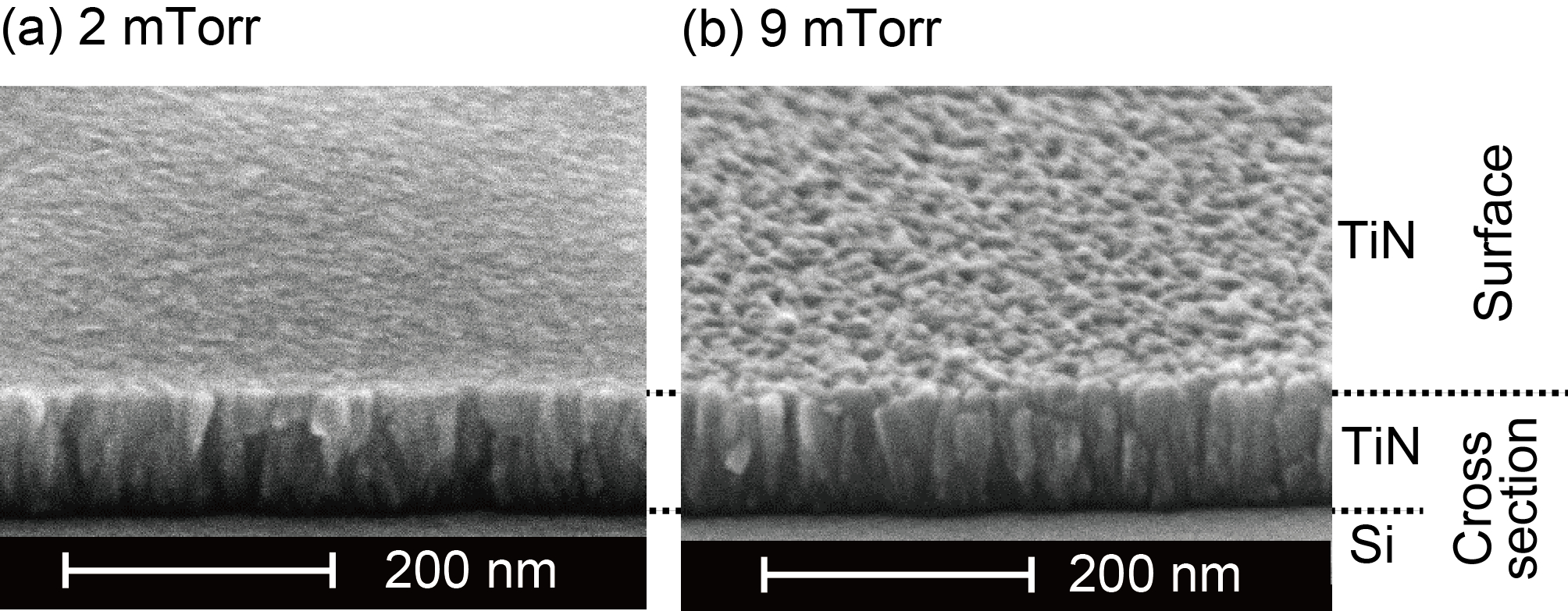}

\end{center}
\caption{(Color online) SEM images of the stoichiometric TiN films deposited at (a) 2 and (b) 9 mTorr. The incident SEM electron-beam direction is tilted by 70$^{\circ}$ from the film-surface normal.}
\label{SEM2-9}
\end{figure}

Figure \,\ref{SEM2-9}(a) and (b) show SEM images of the stoichiometric TiN films for deposition pressures of 2 and 9 mTorr, respectively. The incident electron-beam direction is tilted by 70$^{\circ}$ from the film-surface normal. These images were taken near the center of the substrates (quarter pieces of 3-inch Si wafers). We see that these TiN films are polycrystalline with columnar grains. In the film deposited at 2 mTorr, these grains are intimately bound to their neighbors. In contrast, in the film deposited at 9 mTorr, the grain boundaries are clear, and the surface is rough. The grain boundaries seen in (b) are thought to allow contaminants into the film, which is consistent with the increase in the C + O concentration from $0.3\%$ at 2 mTorr to $13.5\%$ at 9 mTorr, as shown in table \ref{TabContaminations}.

The morphology change due to the pressure increase is similar to the effect of the application of a substrate bias, which has been well studied previously.\cite{Kumar1988} The grains tend to bind as the negative substrate-bias voltage becomes larger. It is known that the atomic peening mechanism leads to a dense structure [as in Fig.\,\ref{SEM2-9}(a)], where the gas atoms reflected from the target with a high momentum pack together the sputtered atoms and increase the density of atoms in each column.\cite{Thornton1985} This effect is more significant at lower pressure, where the reflected atoms have higher momenta. The higher momentum of the sputtered particles at a lower pressure also can help to make such a dense structure.

\begin{figure}
\begin{center}
\includegraphics{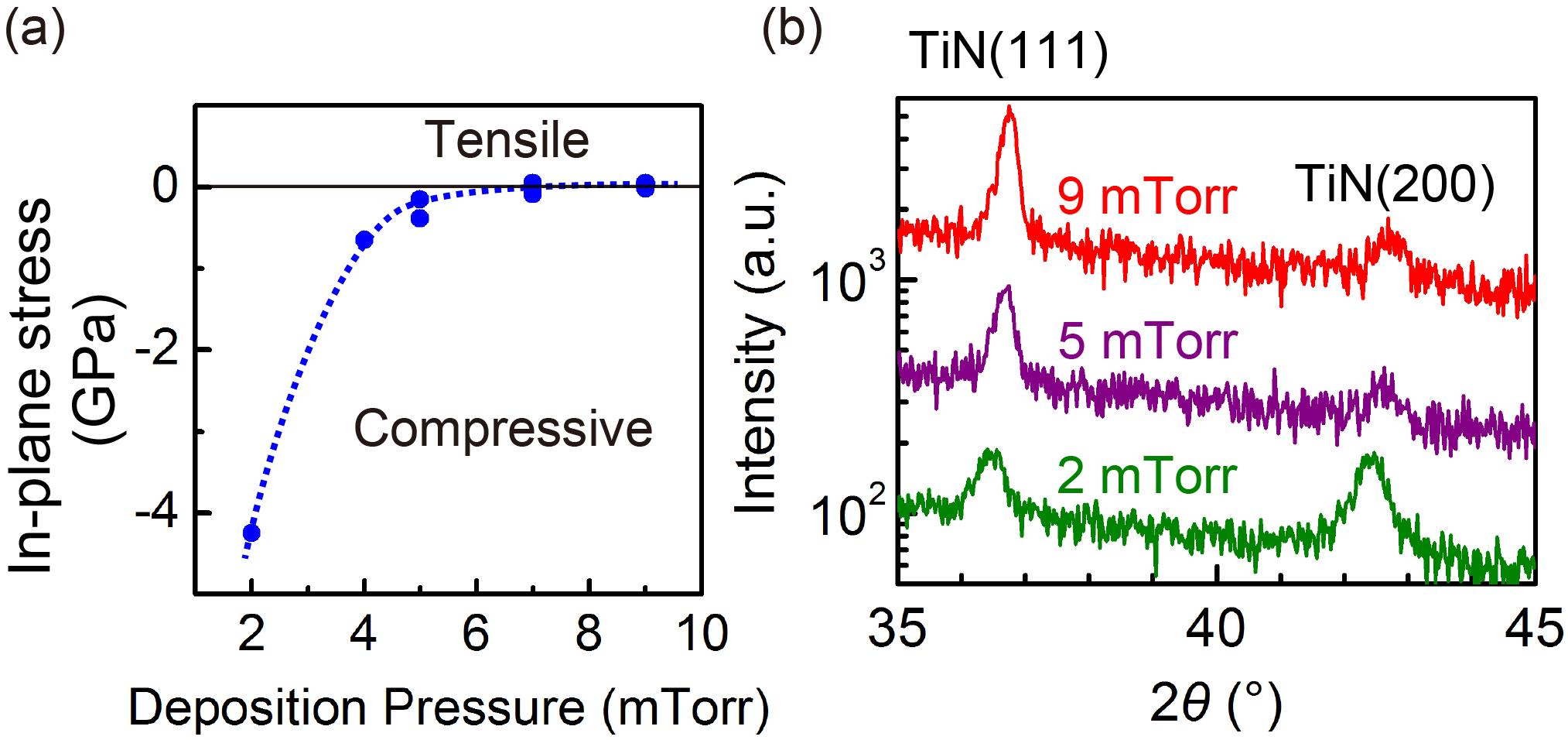}
\end{center}
\caption{(Color online) (a) Stress measured in the stoichiometric TiN films as a function of the deposition pressure. The stress values shown here were measured at the center of 3-inch wafers. (b) The $\omega$-2$\theta$ x-ray diffraction scans of 200-nm-thick stoichiometric TiN films deposited at 2, 5, and 9 mTorr (from bottom to top).}
\label{Stress-XRD}
\end{figure}

The relatively high-atomic-density TiN columns obtained in a low pressure deposition generate a large stress in the film. Figure \,\ref{Stress-XRD}(a) shows the deposition pressure dependence of the in-plane stress in 100-nm thick stoichiometric TiN films. The strain values were determined with a wafer bow measurement (FLX-2320, KLA Tencor, Inc.) at the center of the TiN films deposited on 3-inch Si wafers  at room temperature. When the deposition pressure is low, the film has a strong in-plane compressive strain. By increasing the deposition pressure, the in-plane strain is reduced and changes to a weak tensile strain. This is a very common feature in films deposited by sputtering and is consistent with the SEM results shown in Fig.\,\ref{SEM2-9}.

Figure \,\ref{Stress-XRD}(b) shows x-ray diffraction $\omega$-2$\theta$ scans of the 200-nm-thick stoichiometric TiN films deposited at 2, 5, and 9 mTorr (from bottom to top). In all the spectra, we see two peaks at $\sim$36.5$^{\circ}$ and $\sim$42.5$^{\circ}$ corresponding to the TiN(111) and (200) planes, respectively. The only other peaks detected in our stoichiometric TiN films were from the Si substrate.  With increasing pressure, these TiN peaks shift toward larger angles, and the (111) peak becomes sharper. The lattice constants in the surface-normal direction estimated from these peaks are 0.4260 nm at 2 mTorr and 0.4237 nm at 9 mTorr (corresponding to a 0.5\% decrease in the lattice constant). Since the intrinsic lattice constant of TiN is 0.424 nm, these values are consistent with the tendency of the stress to change as seen in Fig.\,\ref{Stress-XRD}(a).

By increasing the deposition pressure, the dominant crystal orientation changes from (200) to (111). As the mobility of the adatoms is increased by decreasing the deposition pressure, they tend to make closer-packed structures. Thus, the high adatom mobility causes the crystal growth along the (200) orientation, which has the lowest surface free energy.\cite{Patsalas2000}

\begin{figure}
\begin{center}
\includegraphics{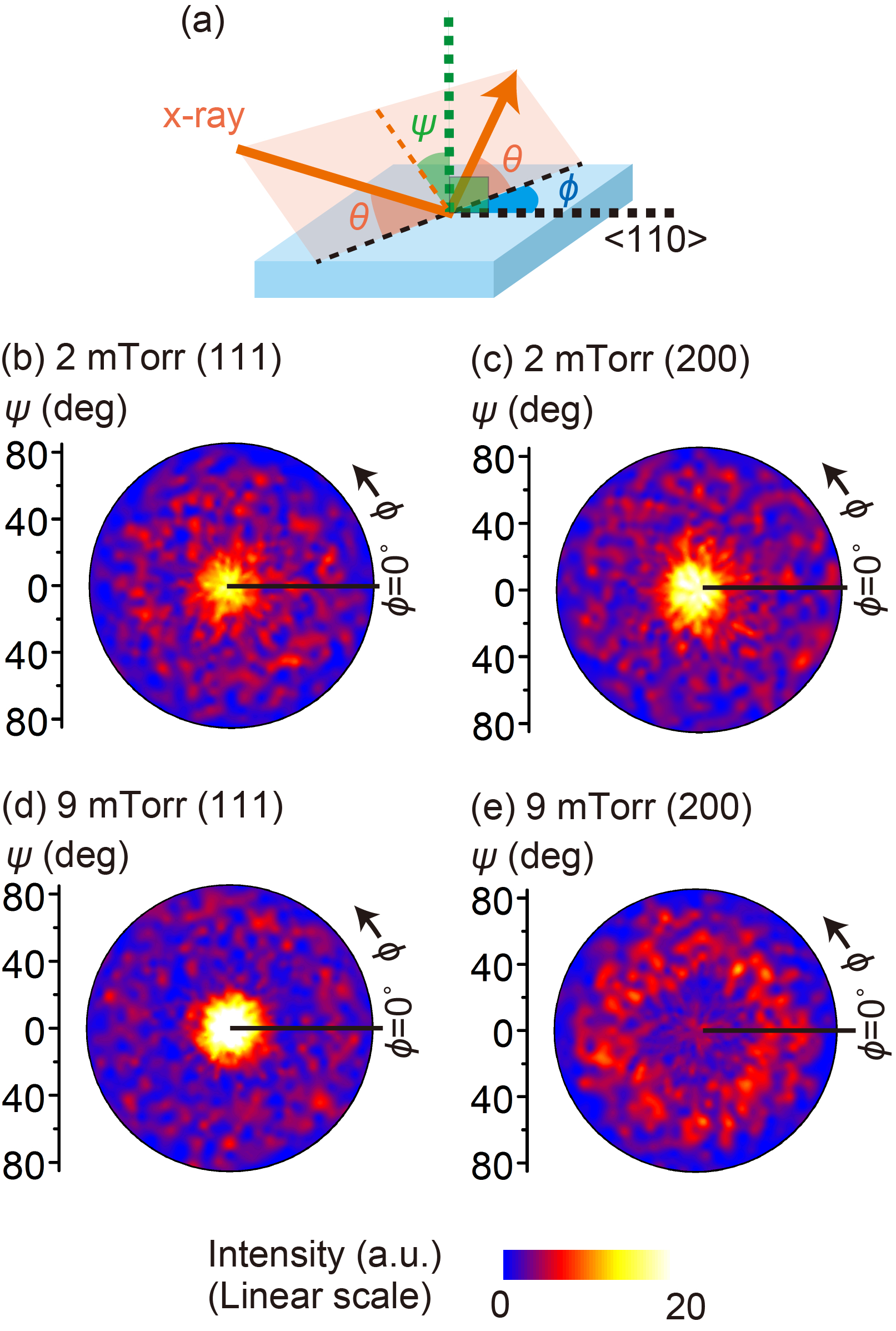}
\end{center}
\caption{(Color) (a) Schematic x-ray beam alignment in our texture measurements. (b), (c) Results of the texture measurements on the stoichiometric TiN film deposited at 2 mTorr when $\theta$ is fixed at (b) 36.66$^{\circ}$ and (c) 42.61$^{\circ}$. (d), (e) The same measurement results when the deposition pressure is 9 mTorr.}
\label{Texture}
\end{figure}

We carried out x-ray texture measurements on the 200-nm-thick stoichiometric TiN films, which give us a more complete understanding of the crystallinity of these films. Figure \,\ref{Texture}(a) shows the schematic x-ray beam alignment in our measurements. We define $\phi$ as the in-plane angle between the Si in-plane <110> axis and the x-ray beam plane [the pink plane in Fig.\,\ref{Texture}(a)], whereas $\psi$ expresses the angle between the measurement direction (broken orange line) and the surface normal of the film. Figure \,\ref{Texture}(b) and (c) show the measurement results of the TiN films deposited at 2 mTorr, when $\theta$ is fixed at 36.66$^{\circ}$ corresponding to TiN(111), and at 42.61$^{\circ}$ corresponding to TiN(200), respectively. We see that the (111) and (200) planes are nearly parallel to the film surface (corresponding to the center yellow zones). The $\psi$ direction of these planes fluctuates slightly ($\psi$=0-20$^{\circ}$). As can be seen in the SEM image of Fig.\,\ref{SEM2-9}(a), the growth directions of the grains are not perfectly aligned in the surface normal direction but have some fluctuation, so these $\psi$ distributions observed in the x-ray texture measurements are probably due to these grains' tilt. In addition, we see that there are randomly oriented (111) and (200) planes (many red areas). Figure \,\ref{Texture}(d) and (e) show the same measurement results on the TiN film deposited at 9 mTorr. In this case, (111) planes nearly parallel to the film surface become dominant. The (200) planes observed at $\psi$=40-70$^{\circ}$ are thought to be the same crystal phase with the (111) plane observed at the center in (d). From these results, we see that the (111) plane is selectively grown in the growth direction in the high pressure condition with a $\psi$ fluctuation up to $\sim$20$^{\circ}$.

\begin{figure}
\begin{center}
\includegraphics{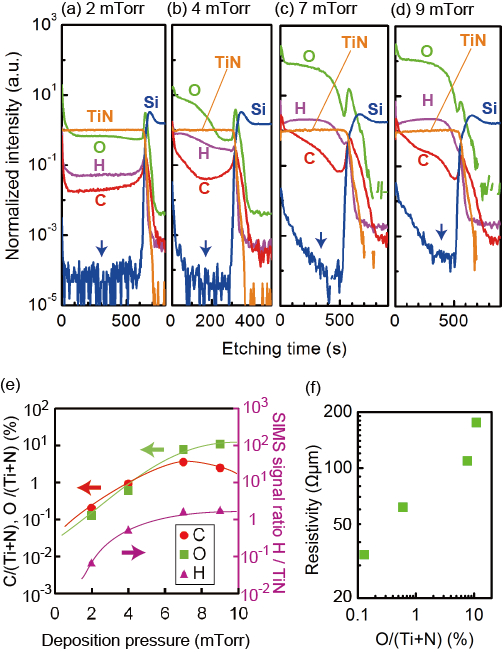}
\end{center}

\caption{(Color) SIMS depth profile of the stoichiometric TiN films deposited at (a) 2 mTorr, (b) 4 mTorr, (c) 7 mTorr, and (d) 9 mTorr as a function of the Ar$^{+}$ etching time from the surface toward the Si substrate. We show the SIMS signals of the H, C, O, Si, and TiN. Here, all the signals are normalized by the intensity of TiN in the TiN-layer region. (e) Deposition pressure dependence of the concentrations of H, C and O. Since it is difficult to determine the H content, we plot the integrated SIMS intensity of H over the thickness divided by that of TiN. The actual concentrations of C and O were estimated from the total SIMS signals in (a)-(d) using the RBS data of the TiN film deposited at 9 mTorr as a reference.  (f)  The dependence of resistivity on O content for the nearly stoichiometric TiN films.}
\label{SIMS2-9}
\end{figure}
\begin{table}
\caption{Ar and N$_{2}$ flow rates used to deposit nearly stoichiometric TiN films, and the concentrations of C and O relative to the sum of the Ti and N content. The C and O concentrations were estimated from SIMS intensities integrated over the film thickness, using RBS data of the TiN film deposited at  9 mTorr as a reference.}
\begin{tabular}{|c|c|c||c|c|}
\hline
\hline
Pressure & \multicolumn{2}{|c||}{Flow rate (sccm)} & \multicolumn{2}{|c|}{Concentration (\%)} \\
\cline{2-5}
 (mTorr) & \ \ \ Ar \ \ \ & N$_{2}$ & \ \ \ \ C \ \ \ \ & O \\
\hline
\hline
2 & 15 & 3.5 & 0.2 & 0.1 \\
\hline
4 & 15 & 3.5 & 0.9 & 0.6 \\
\hline
7 & 15 & 2.5 & 3.6 & 8 \\
\hline
9 & 15 & 2.25 & 2.5 & 11 \\
\hline
\hline
\end{tabular}
\label{TabContaminations}
\end{table}

Figure \,\ref{SIMS2-9}(a)-(d) shows the secondary ion mass spectroscopy (SIMS) results from the TiN films deposited at 2, 4, 7, and 9 mTorr. The thicknesses of these films are 200, 100, 100, and 100 nm, respectively. These TiN films were deposited on quarter pieces of 3-inch Si substrates. We etched the film from the surface toward the Si substrate with different etching rates depending on the sample. The Si interface is identified by the increase in the Si signal. Here, all the signals are normalized by the intensity of TiN in the TiN-layer region in each graph. Thus, we can directly compare the relative concentrations between the samples. We note that the low level signals of Si in the TiN layer (see the arrows) are not coming from Si, but probably coming from a contaminant that has the same mass number (29) as is used to detect Si, such as COH. The concentrations of H, C, and O strongly increase with increasing deposition pressure. The TiN films incorporate the contaminants when they are exposed to air after deposition.\cite{Logothetidis1999} We have observed this directly by witnessing the films change color from the golden color of stoichiometric TiN to the green color of titanium oxynitride as the load lock is being vented. 

From our RBS measurements on the 100-nm thick TiN film deposited at 9 mTorr, C and O contents relative to the sum of the Ti and N contents are estimated to be 2.5\% and 11\%, respectively. Using these values and integrating the SIMS intensities shown in Fig.\,\ref{SIMS2-9}(a)-(d), we can estimate the contamination levels of C and O as shown in Table \ref{TabContaminations}. The deposition pressure dependence of these concentrations is shown in Fig.\,\ref{SIMS2-9}(e). As shown in Fig. 1, the resistivity increases and $T{_\text{c}}$ decreases as the deposition pressure is increased; this change is likely due to rising contaminant levels in the TiN.  In Fig. \ref{SIMS2-9}(f) we show the strong dependence of resistivity on oxygen content, indicating that the regulation of oxygen content is essential to the control of the electronic properties of the film.

As shown in Fig.\,\ref{SEM2-9}, the grain boundaries become more defined as the deposition pressure increases, and the surface area, which can absorb the contaminants, becomes larger as a consequence. It has been reported that x-ray photoemission spectroscopy measurements on sputtered TiN films show a Ti 2$p{_3}$${_\slash}$${_2}$ peak at 458 eV which corresponds to TiO$_{2}$,\cite{Logothetidis1999} which suggests that some of the O atoms may enter the TiN crystal and react with Ti atoms. Under high pressure conditions, bonds between Ti and N become weak due to the low kinetic energy of the sputtered particles, which is also thought to be the origin of such a high level of contaminants in the TiN films deposited at high pressure.

\section{IN-PLANE DISTRIBUTION}
\label{IN-PLANE DISTRIBUTION}

\begin{figure}
\begin{center}
\includegraphics{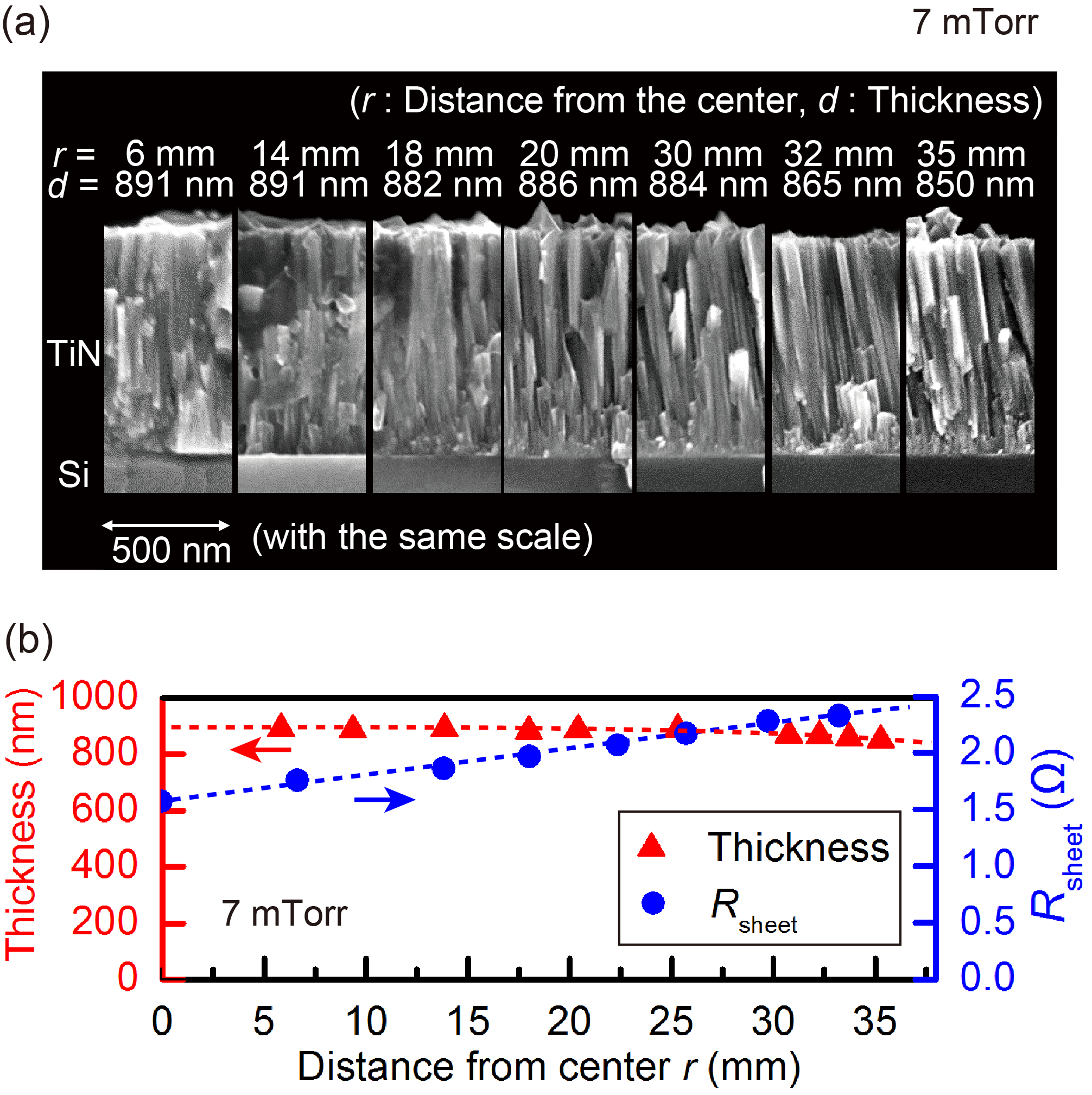}
\end{center}

\caption{(Color online) (a) Cross-sectional SEM images of the stoichiometric TiN film deposited at 7 mTorr on a 3-inch Si(001) wafer. Here, we show the images taken at positions from 6 to 35 mm measured from the center of the wafer. (b) $R$$_{\text{sheet}}$ (red triangles) and film thickness (blue circles) as a function of the distance from the center of the wafer.}
\label{InPlaneDist}
\end{figure}

As shown above, the TiN film properties directly depend on the kinetic energy of the sputtered particles, which can be controlled by deposition pressure. In magnetron sputtering, the in-plane energy distribution of the sputtered particles is inevitable because there is an in-plane inhomogeneity of the plasma intensity caused by the spatial variation of the magnetic field from the magnetron. Figure \,\ref{InPlaneDist}(a) shows the cross-sectional SEM images of a stoichiometric TiN film deposited on a 3-inch Si(001) wafer at 7 mTorr. We show the images taken at 6-35 mm from the center of the wafer. There is a similar tendency seen in the pressure dependence; the grain boundary becomes clearer from the center to the edge. In Fig. 6(b), we show the measured sheet resistance $R$$_{\text{sheet}}$ (blue circles) and the film thickness (red triangles) as a function of the distance from the center of the wafer. We see that the film thickness is reduced by 5\% from 890 to 850 nm, whereas, $R$$_{\text{sheet}}$ increases by 70\% from 1.5 to 2.5 $\Omega$. This large increase in $R$$_{\text{sheet}}$ cannot be explained by the 5\% thickness reduction alone.

\begin{figure}
\begin{center}
\includegraphics{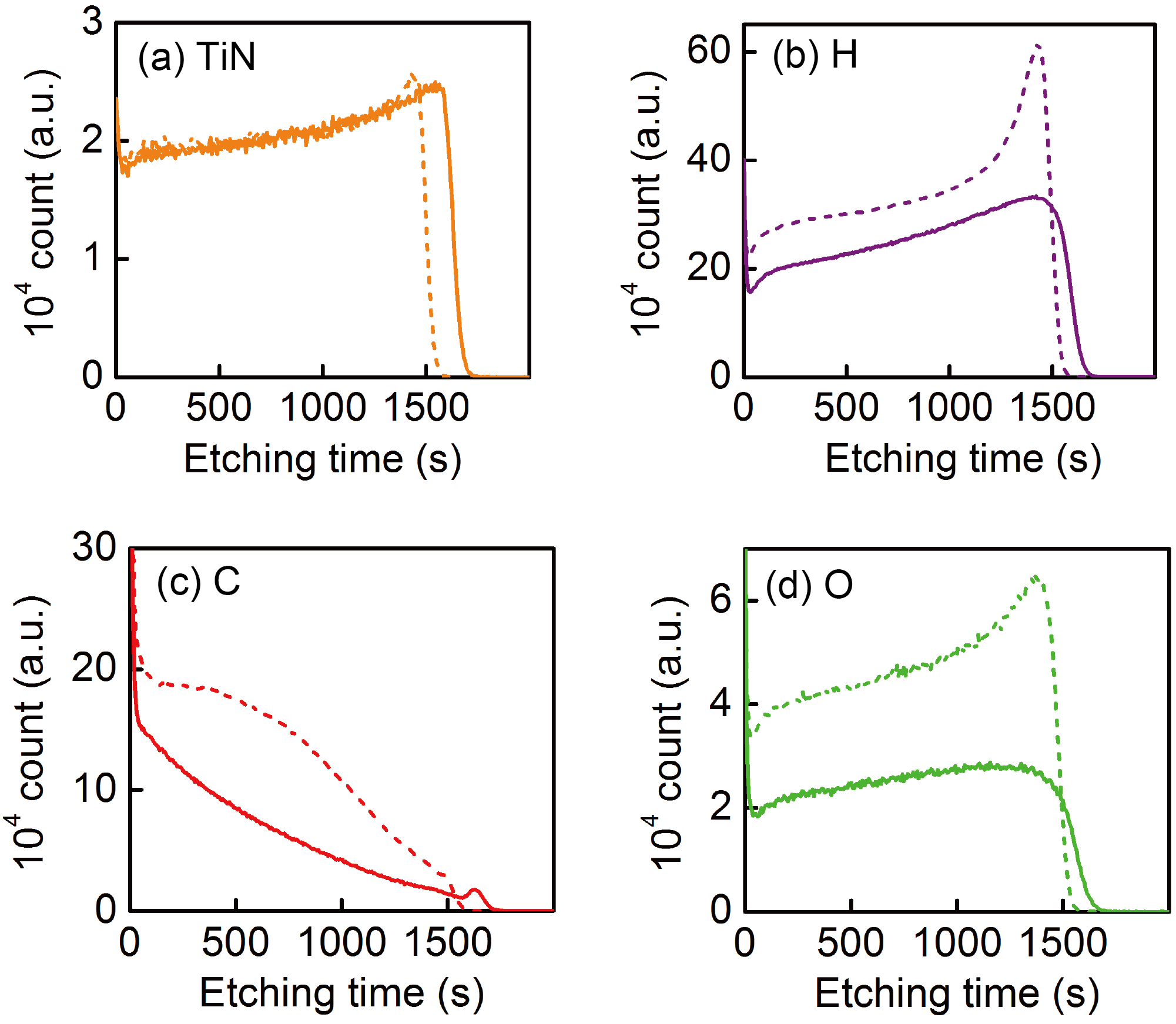}
\end{center}
\caption{(Color online) SIMS depth profiles of (a) TiN, (b) H, (c) C, and (d) O obtained in the stoichiometric TiN film deposited at 7 mTorr. The solid curves are those at the center, and the dotted curves are the ones at the edge of the 3-inch wafer.}
\label{InPlaneDist-SIMS}
\end{figure}

Figure \,\ref{InPlaneDist-SIMS} (a)-(d) shows the SIMS depth profiles of TiN, H, C, and O on the stoichiometric TiN film deposited at 7 mTorr. The solid curves are the SIMS intensities at the center, and the dotted curves are the ones at the edge of the 3-inch wafer. Here, we show the raw data. We see that the TiN contents are the same between at the center and at the edge. However, all the contaminant levels are higher at the edge than those at the center.  This suggests that the in-plane energy distribution of the sputtered particles causes the large distribution of the resistance as well as the contamination.

\section{T-S DISTANCE DEPENDENCE}
\label{T-S DISTANCE DEPENDENCE}
\begin{figure}
\begin{center}
\includegraphics{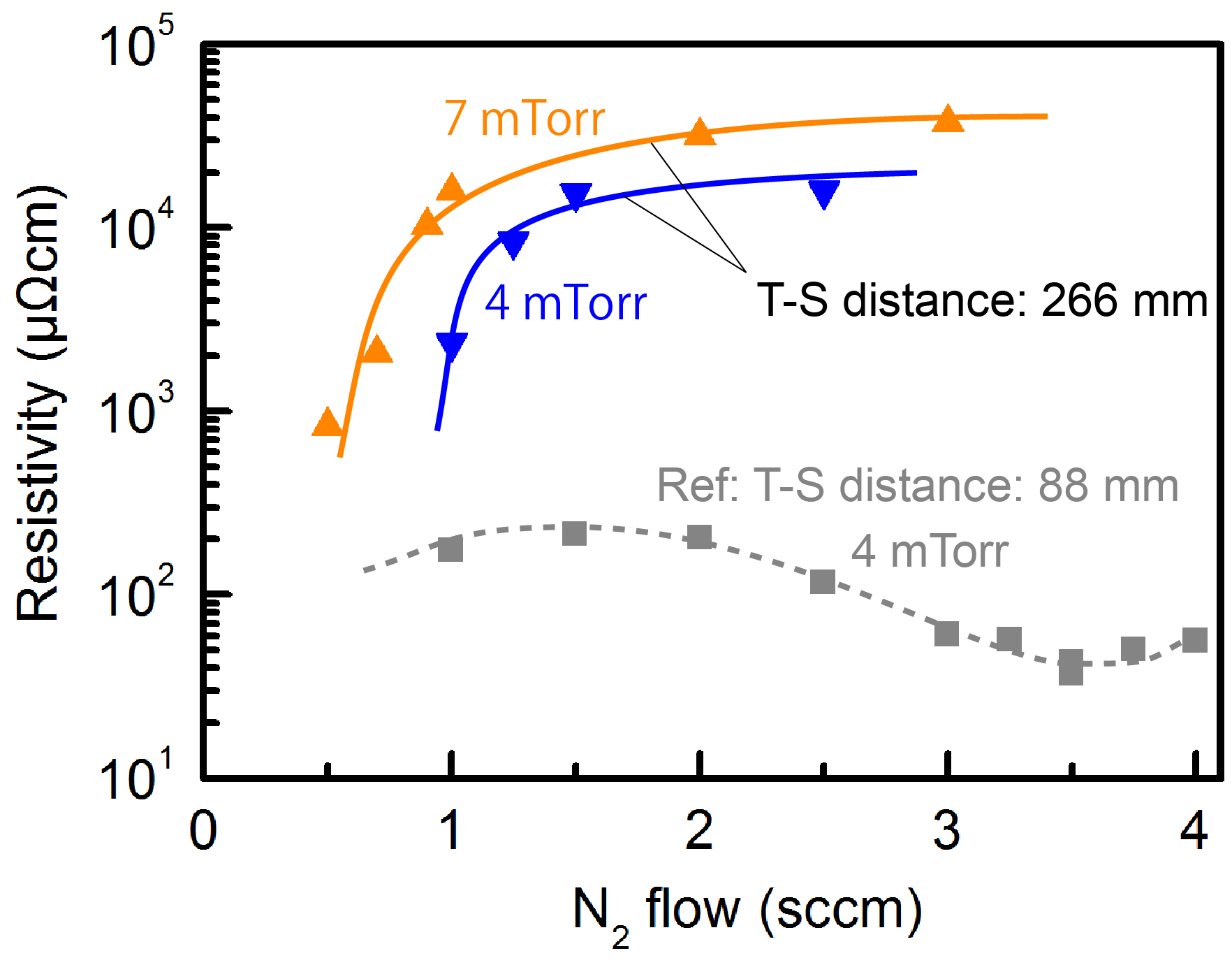}
\end{center}
\caption{(Color online) The orange (triangle) and blue (inverse triangle) points are the resistivities of the TiN films as a function of the N$_{2}$ flow rate when the T-S distance is 266 mm and the deposition pressure is 7 and 4 mTorr, respectively. As a reference, we show data for films deposited at 4 mTorr with the T-S distance at 88 mm (gray rectangles), which is the same T-S distance used for all the depositions of the TiN films shown in sections \ref{BASIC PROPERTIES} through \ref{IN-PLANE DISTRIBUTION}. The solid and the broken curves are guides for eyes.}
\label{TS-change}
\end{figure}

The energy of the sputtered particles just before reaching the film surface also depends on the T-S distance. The sputtered particles experience more collisions for larger T-S distances. Here, we studied the effect of the T-S distance on film quality. The orange (triangle) and blue (inverse triangle) points in Fig.\,\ref{TS-change} show the resistivities of the TiN films as a function of the N$_{2}$ flow rate when the T-S distance was 266 mm and the deposition pressure was 7 and 4 mTorr, respectively. As a reference, we show the films deposited at 4 mTorr with a T-S distance of 88 mm (the same T-S distance as was used for all depositions of the TiN films shown in the previous sections). The resistivity becomes two orders of magnitude higher as the T-S distance is changed from 88 to 266 mm. The films deposited with a large T-S distance (288 mm) have a green color, which is the typical color of titanium oxynitride (TiNO). By carrying out RBS analysis on the film deposited at 7 mTorr with a N$_{2}$ flow rate of 3 sccm and a T-S distance of 266 mm, the carbon and oxygen contents are estimated to be 5\% and 27\%, respectively. We found that the color of these films was gold just after the deposition when they are still in the vacuum chamber. However, while venting the load lock with N$_{2}$ gas (purity: 99.999\%) with TiN samples inside, the film color was observed to change from gold to green.  This shows that the grain boundary surfaces of the TiN films are highly reactive since H, C, and O were gettered from the nearly pure N$_{2}$. This result indicates that TiN films deposited with low sputtered-energy particles absorb a high amount of these contaminants, and the contaminant levels strongly depend on the sputtered particle energy. Therefore, the T-S distance is important for the production of high-quality TiN films.\cite{T-SComment}
\section{PROPERTIES OF THE SCPW RESONATORS}
The application of superconducting films to circuits that detect photons and process quantum information motivates our investigation of the relationships between deposition conditions and the properties of the obtained material. Coupling this knowledge with an understanding of the impact that these materials properties have on device performance enables the realization of thin films optimized for quantum circuits.  Whereas in the previous sections we discussed the response of TiN thin films to the reactive sputtering parameters used in their production, here we inspect the influence of specific TiN film features on the performance characteristics of devices into which the films are made.  In order to evaluate these effects, we have deposited a series of nearly stoichiometric thin films at several pressures (2,4,5, and 7 mTorr) using the procedure outlined in the preceding sections with a T-S distance of 88 mm.  Next, we patterned these films into SCPW microwave resonators and compared the performance of these devices to the material properties of either an unprocessed section of the same sample or of a companion wafer deposited under nominally identical conditions.  The resonators in our experiment take the form of a quarter wavelength segment of coplanar waveguide terminated at opposite ends by an electrical open circuit and short circuit to the ground plane.

Among the virtues of the SCPW resonator is the simplicity of its fabrication.  The SCPW resonator thus provides a context for our study of material vs device performance that is protected against the conflation of fundamental material data and complications arising from involved cleanroom proceedings.  In this experiment, the substrate preparation and deposition described previously were followed by a single optical lithography and etching sequence. The nominally 100 nm TiN films were etched in an inductively-coupled plasma etcher with Cl$_{2} $ as the reactive species, under conditions yielding an etch rate of approximately 5.5 nm/sec.  This primary etch was followed with a secondary, 5 sec SF$_{6}$ etch, as this has been shown to reduce loss \cite{Sandberg2012}.  This reliably produced a SCPW structure with a silicon substrate trench depth of 56 $\pm$ 11 nm.  Once etched, the wafers were diced to yield chips measuring 6.25 x 6.25 mm square.  Dies drawn from the center regions of our wafers were packaged in an Al sample box with an approximate linear wirebond density of 3/mm from the box to the device ground plane, in preparation for measurement.

During measurement, the sample box was mounted on the cold plate of an adiabatic demagnetization refrigerator with a base temperature of $\sim$ 70mK.  This cold plate is shielded in stages from infrared radiation and external magnetic fields.\cite{Barends2011}The output of the test chip was connected to a high-electron-mobility transistor amplifier at 4K, followed by a room temperature amplifier chain. A vector network analyzer (Agilent 5230A or 5242A) was used for excitation and detection.  We characterized the SCPW by measuring the transmission scattering parameter S$_{21}$ of a microwave transmission line capacitively coupled to a resonator.  Details of the measurement and analysis are supplied by Megrant \textit{et al}. \cite{Megrant2012}

\begin{figure}
\begin{center}
\includegraphics{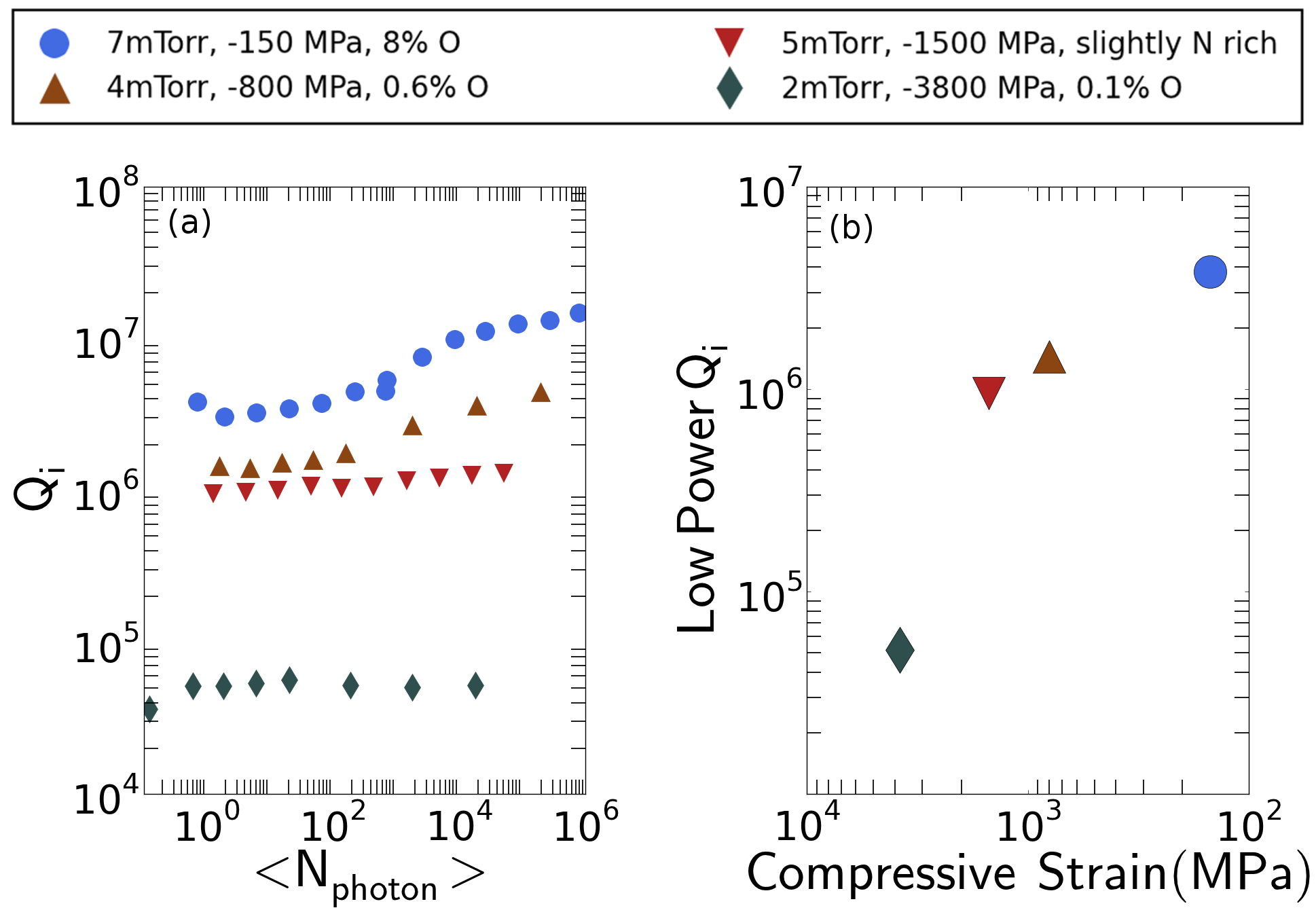}
\end{center}

\caption{(Color online) (a) The power dependence of the SCPW resonator $Q_{i}$ for the nearly stoichiometric films deposited at 2,4,5, and 7 mTorr and respective strain values -3800, -800, -1500, and -150 M Pa, expressed in terms of the expectation value of the resonator's photonic occupation number.  (b) The strain dependence of the low power $Q_{i}$ measured near $\langle N_{photon}\rangle =1$.}
\label{QivsNp}
\end{figure}

 Because energy relaxation events form an important error class in quantum information processing, this study focused on maximizing the low power $Q_{i}$, which is generally agreed to be limited by coupling to TLSs in the low energy excitation regime that is relevant to quantum computing and accessed in the limit where the number of photons in the resonator is small.  Figure \ref{QivsNp} (a) reports the dependence of $Q_{i}$ on the microwave drive power for devices made from the nearly stoichiometric films deposited at 2,4,5, and 7 mTorr with N$_{2}$ flow rates of 3.5, 3.5, 3.5, and 2.5 sccm, which had compressive film strain values of  3800, 800, 1500, and 150 MPa, respectively.  The 5mTorr sample was deposited under slightly nitrogen-rich conditions and we believe that this explains the nonmonotonicity of film strain vs deposition pressure for these samples.  Figure \ref{QivsNp} (b) shows the measured quality factors, at resonator excitation energies approximately equal to that of a single photon at the resonance frequency, versus the strain of the film from which they were produced.  For Fig. 9 (a) and (b) we have selected the best performing device from each sample to establish an upper bound on device performance for a  given set of material parameters, independent of such systematic issues as cleanroom process variability \cite{qiplotfn} and time dependent external magnetic fields.  We note that $Q_{i}$ is enhanced as the film strain decreases and the oxygen content increases.  In particular, the lowest strain film produced a resonator with a measured low power $Q_{i} = 3.8 \times 10^{6}$, the highest reported value to date; however,  a subsequent reproduction attempt with a nominally identical film was unsuccessful.  Aside from that, we find that a respectable low power $Q_{i} >1 \times 10^{6}$ can be reliably achieved with low strain TiN SCPW resonators.
 
Since surface oxides of superconducting electrodes and their substrates are supposed to harbor the TLS populations\cite{Gao2008} responsible for limiting the low power quality factors of SCPWs, the result that increased oxygen concentration in our compound superconducting films is associated with an increase in quality factor was not expected.  More surprising, perhaps, is the absolute magnitude of the oxygen concentration; in the best performing films the ratio of O to TiN was measured to be 8$\%$.  We interpret this result to suggest that the presence of oxygen is not, in and of itself, deleterious to the resonator quality factor.  Rather, oxygen incorporation by the TiN crystal may be an innocuous byproduct of the low strain condition which we suspect is itself fundamentally responsible for the quality factor increase in our experiment.

Although significant oxygen impurity concentrations in our TiN films do not directly translate to increased loss in our microwave resonators, the presence of oxygen in the TiN crystal is not inconsequential.  As seen in Fig. 5(f), the normal state resistivity of our films depends strongly on the oxygen content.  From a SCPW resonator perspective, the salient consequence of increased resistance is a resonant frequency reduction via kinetic inductance augmentation.  This is a large effect in TiN, where the kinetic inductance is often larger than the geometric inductance for common device geometries.  Applying these facts to the uniformity data in Fig.\,\ref{InPlaneDist}(b)  presents an engineering dilemma to those who would make quantum integrated circuits from reactively sputtered TiN: in this material system, oxygen may not function as an instrument of excess loss, but the uncontrolled manner by which it installs itself in the crystal, and the corresponding lack of uniformity, renders engineering large-scale circuits on reactively sputtered TiN difficult at this time.								

\section{SUMMARY}
We have shown a detailed picture of the properties of TiN films deposited at room temperature by varying the deposition pressure and the N$_{2}$ flow rate. When fixing the deposition pressure, the resistivity minimum corresponds to the stoichiometric point ($x$=0.5). By increasing the deposition pressure, while keeping $x$=0.5, the resistivity rises and $T_{\text{c}}$ decreases. The strong in-plane compressive stress changes to weak tensile stress as the deposition pressure increases. The dominant crystal orientation changes from (200) to (111). The grain boundaries become clearer, and the contamination levels, including H, C, and O, significantly increase. The grain boundaries play a crucial role in the absorption of the contaminants. This morphology change is thought to be induced by the energy change of the sputtered particles due to the change of the deposition pressure.

The in-plane particle energy distribution caused by the in-plane inhomogeneity of the plasma leads to a large radial resistivity change (70 \%) across a TiN film deposited at 7 mTorr on a 3-inch wafer. From the center to the edge of this sample, the grain boundaries become clearer, which is very similar to the effect of deposition pressure increase. We have found that larger amounts of the contaminants H, C, and O exist at the edge of the wafer than at the center. By increasing the T-S distance from 88 mm to 266 mm, the film color was changed from gold to green, and we detected a higher amount of H, C, and O in the film deposited with the T-S distance at 266 mm than at 88 mm. The energy of the sputtered particles, which decreases with distance from the substrate center and with increasing the T-S distance, is also responsible for these phenomena.

Following the method developed in section III of this paper, we deposited nearly stoichiometric films at 2,4,5, and 7 mTorr and found that increasing the deposition pressure decreased the film strain and increased the oxygen content.  These changes were associated with an increase in the Q$_{i}$ factor of SCPW resonators made from these films.  However, the film resistivity is a strong function of oxygen content and was found to vary considerably with distance from the center of the substrate.  The variation in resistivity, which translates to a surface inductance variation in the superconductor, makes it difficult to design and produce large circuits without addressing this issue first.  However, the resonator performance achieved with TiN implies that this material is immediately useful for smaller circuits and justifies efforts to engineer more uniform deposition methods for larger devices.

\section*{ACKNOWLEDGEMENTS}%

We thank M. R. Vissers, H. Sukegawa, and H. G. Leduc for the technical advice regarding the sputtering of TiN. S. O. acknowledges the Japan Society for the Promotion of Sciences (JSPS) for a Postdoctoral Fellowship for Research Abroad.  Devices were made at the UC Santa Barbara Nanofabrication Facility, a part of the NSF-funded National Nanotechnology Infrastructure Network. This research was funded by the Office of the Director of National Intelligence (ODNI), Intelligence Advanced Research Projects Activity (IARPA), through Army Research Office Grant No. W911NF-09-1-0375. All statements of fact, opinion or conclusions contained herein are those of the authors and should not be construed as representing the official views or policies of IARPA, the ODNI, or the U.S. Government.


\begin{thebibliography}{99}
\bibitem{qiplotfn} Following the resonator measurements, several devices were found to have some residual photo resist from fabrication.  Since the best performing device was observed to have resist remaining after the striping procedure and only the best device from each film is reported here, it is unlikely that this significantly limited the stated performance of any film.
\end{thebibliography}
\end{document}